\documentclass{aa}
\usepackage{graphicx,times,amssymb}
\usepackage[english]{babel}
\renewcommand{\d}{\mathrm{d}}

\begin{document}

\title{Measuring $\Omega_0$ with higher-order Quasar-Galaxy
  Correlations induced by Weak Lensing}
\author{Brice M\'enard\inst{1,2}, Matthias Bartelmann\inst{1} \and
  Yannick Mellier\inst{2,3}}
\institute{$^1$Max-Planck-Institut f\"ur Astrophysik, P.O.~Box 1317,
  D--85741 Garching, Germany\\
  $^2$Institut d'Astrophysique de Paris, 98 bis Bld Arago, F--75014,
  Paris, France\\
  $^3$LERMA, Observatoire de Paris, 61 avenue de l'Observatoire,
  F--75014 Paris, France}

\date{\today}

\authorrunning{M\'enard et al.}
\titlerunning{Higher-order QSO-Galaxy Correlations}

\abstract
  { Via the magnification bias, gravitational lensing by large-scale
   structures causes angular cross-correlations between distant quasars
   and foreground galaxies on angular scales of arc minutes and
   above. We investigate the three-point cross-correlation between
   quasars and galaxy pairs measurable via the second moment of the
   galaxy counts around quasars and show that it reaches the level of
   a few per cent on angular scales near one arc minute. Combining
   two- and three-point correlations, a skewness parameter can be
   defined which is shown to be virtually independent on the shape and
   normalisation of the dark-matter power spectrum. If the galaxy bias
   is linear and deterministic, the skewness depends on the cosmic
   matter density parameter $\Omega_0$ only; otherwise, it can be used
   to probe the linearity and stochasticity of the bias. We
   finally estimate the signal-to-noise ratio of a skewness
   determination and find that around twenty thousand distant quasars e.g.~from the
   Sloan Digital Sky Survey should suffice for a direct measurement of
   $\Omega_0$.
  }

\maketitle

\section{Introduction}

It is widely believed that structures and galaxies in the Universe formed  from 
gravitational growth of Gaussian primordial mass density  fluctuations dominated 
 by dark matter. Direct support for this picture is provided by the recent weak
lensing surveys of galaxies which measured  the 
systematic distortion of faint background-galaxy images produced by the 
gravitational tidal field of intervening dark-matter inhomogeneities~: 
the \emph{cosmic shear}
 (Bacon et al. 2000 and 2002; H\"ammerle et al.;
2002; Hoekstra et al. 2002; Kaiser et al. 2000; Maoli et al. 2001;
 R\'efr\'egier et al 2002;
Rhodes et al. 2001; Van Waerbeke et al. 2000, 2001, 2002; Wittman et
al. 2000).   The shape of the cosmic shear signal as a function of angular 
 scale remarkably follows theoretical expectations, which successfully 
  confirms the gravitational instability scenario, even on small scales 
where non-linear structures dominate the lensing signal.  

Further evidence for lensing is provided by the gravitational \emph{magnification bias}. 
In addition to distortion, distant objets are magnified or demagnified, 
depending on whether the matter along their
lines-of-sight is over- or underdense compared to the mean. Magnified
sources are preferentially included into flux-limited samples, thus
sources behind matter overdensities are somewhat
over-represented. Since galaxies are biased with respect to the
dark-matter distribution, it is expected that this effect induces cross-correlations
between distant sources and foreground galaxies.  The 
existence of significant cross-correlations between distant
quasars and foreground galaxies on angular scales of several arc
minutes has indeed been firmly established (see Bartelmann \& Schneider 2001
for a review) and motivated further theoretical development in 
order to predict how the magnification bias depends on cosmological 
models. Following earlier work by Bartelmann (1995) and 
Dolag \& Bartelmann (1997), M\'enard \& Bartelmann (2002) demonstrated the
high sensitivity of  angular quasar-galaxy cross-correlation
    function to several cosmological parameters, namely the
matter density parameter, $\Omega_0$, the normalisation and shape of
the dark-matter power spectrum, $\sigma_8$ and $\Gamma$, and the bias
parameter of the galaxies, $b$. Hence, magnification bias of quasars is
 equally  efficient as cosmic
shear in constraining the geometry and the dark matter power spectrum of the
Universe.  However, as with cosmic shear, the information provided by the
quasar-galaxy correlation function alone is insufficient for independently
constraining all these parameters. 

Following similar motivations as Bernardeau, van Waerbeke \& Mellier 
(1997) and Jain \& Seljak (1997), we decided to explore how 
  deviations from Gaussian statistics produced by non-linear growth of structures 
could modify the galaxy-quasar cross-correlation signal and eventually 
break some degeneracies between comological parameters.
   The easiest approach is to focus on the additional information that can be
 extracted from higher-order correlations between quasars and galaxies
 which are most sensitive to non-Gaussianity,
 namely the correlation between distant quasars and foreground
galaxy \emph{pairs}. 
 As for the skewness of the convergence field, we can expect that some
parameter dependencies disappear by normalising the 
three- with two-point correlations.

The paper is structured as follows. We briefly present the formalism
of the quasar-galaxy correlation function in Sect.~\ref{basics}, assuming
 the paradigm of gravitational instability of a Gaussian random field is valid.
  In Sect.~\ref{3-point}, we then introduce the quasar-galaxy-galaxy
correlator and predict some useful observational
signatures. Section~\ref{evaluation} deals with density statistics and
the numerical evaluation of the triple correlator. We then define a
skewness parameter in Sect.~\ref{mesuring_omega} and demonstrate how
it can be used for directly measuring $\Omega_0$. Similarly, we show
in Sect.~\ref{mesuring_omega} how several properties of the galaxy bias
can be constrained. We finally estimate the signal-to-noise ratio of
the corresponding observation, and specialise it for the \emph{Sloan
Digital Sky Survey} in Sect.~\ref{section_sn}.

\section{Magnification-induced correlation functions}
\label{basics}

Basic statistical properties of the magnification due to gravitational
lensing by large-scale structure have been studied in earlier
papers. To lowest order in the relevant quantities, magnification is
proportional to the lensing convergence, which has identical
statistical properties to the lensing-induced distortions, i.e.~the
cosmic shear. Those were investigated in many studies, starting with
the pioneering papers of Gunn (1967) and Blandford et al.~(1992).

The two-point correlation function caused by gravitational
magnification between background quasars and foreground galaxies was
first introduced by Bartelmann (1995) and generalised by Dolag \&
Bartelmann (1997). We refer the reader to these papers for detail and
only briefly recall the formalism and approximations leading to the
two-point correlation function. The notation and the definitions
used for the three-point correlation function are presented in the
next section.

\subsection{The two-point correlation function $w(\theta)$}

The angular two-point correlation function between quasars and
galaxies is defined by
\begin{equation}
  w_\mathrm{QG}(\theta) = \frac
   {\left\langle
    [n_\mathrm{Q}(\vec\phi)-\bar{n}_\mathrm{Q}]\,
    [n_\mathrm{G}(\vec\theta+\vec\phi)-\bar{n}_\mathrm{G}]
    \right\rangle}
   {\bar{n}_\mathrm{Q}\,\bar{n}_\mathrm{G}}\;,
\label{eq_main}
\end{equation}
where $n_\mathrm{Q,G}$ are the number densities of quasars and
galaxies on the sky. The bar denotes the average on the sky, and the
angular brackets denote averaging over positions $\vec\phi$ and the
directions of $\vec\theta$, assuming isotropy.

Lensing magnification increases the flux received from sources behind
matter overdensities, but also stretches the sky and thus dilutes the
sources, modifying their number density. The net effect, an increase or a
 decrease of the source number density, is called the
magnification bias. It depends on the number of sources gained per
solid angle by the flux magnification. Let $\alpha$ be the logarithmic
slope of the source number counts as a function of flux, then the
number-density fluctuation is
\begin{equation}
  \frac{n_\mathrm{Q}(\vec\theta)-\langle n_\mathrm{Q}\rangle}
       {\langle n_\mathrm{Q}\rangle}=
  \delta\mu^{\alpha-1}(\vec\theta)\;,
\end{equation}
where $\delta\mu(\vec\theta)=\mu(\vec\theta)-1$ is the magnification
fluctuation as a function of position $\vec\theta$. The magnification
is related to convergence $\kappa$ and (the complex) shear $\gamma$ by
\begin{equation}
  \mu=\left[(1-\kappa)^2-|\gamma|^2\right]^{-1}\;.
\end{equation}
Thus, to first order in $\kappa$ and $\gamma$, the magnification
fluctuation is
\begin{equation}
  \delta\mu\equiv\mu-1=2\kappa\;.
\end{equation}

Assuming that galaxies are linearly biased with respect to the
dark-matter distribution, we can write
\begin{equation}
  \frac{n_\mathrm{G}(\vec\theta)-\langle n_\mathrm{G}\rangle}
       {\langle n_\mathrm{G}\rangle}=
  \bar b\,\bar\delta(\vec\theta)\;,
\end{equation}
where $\bar b$ is the averaged bias factor of all considered galaxies, 
and $\bar\delta$ is the projection
\begin{equation}
  \bar\delta(\vec\theta)=\int_0^{w_\mathrm{H}}\d w\,
    p_\delta(w)\,\delta[f_K(w)\vec\theta,w]\;,
\label{eq:5}
\end{equation}
of the density contrast $\delta$ along the line-of-sight, weighted by
the normalised distance distribution $p_\delta(w)$ of the observed
galaxies which are cross-correlated with the quasars. Here, $w$ is the
comoving radial distance along the line-of-sight, $f_K(w)$ is the
comoving angular diameter distance, and the upper integration boundary
$w_\mathrm{H}$ is the comoving radial distance to the ``horizon'' at
$z\to\infty$ (For more sophisticated biasing models, e.g.~including
stochasticity or non-linearity, see Pen 1998; Dekel \& Lahav 1999; 
Somerville et al. 2001.)

We can thus write the two-point correlation function introduced in
Eq.~(\ref{eq_main}) as
\begin{equation}
  w_\mathrm{QG}(\vec\theta)=\bar b\,\langle
    \mu^{\alpha-1}(\vec\phi)\,\bar\delta(\vec\phi+\vec\theta)
  \rangle\;.
\end{equation}
First-order Taylor expansion of the magnification around unity yields
\begin{equation}
  w_\mathrm{QG}(\vec\theta)=2\,\bar b\,(\alpha-1)\langle
    \kappa(\vec\phi)\,\bar\delta(\vec\phi+\vec\theta)
  \rangle\;.
\label{wqg}
\end{equation}
Like $\bar\delta$, $\kappa$ is a weighted projection of the density
contrast along the line-of-sight. Specifically, we can write $\kappa$
as
\begin{equation}
  \kappa(\vec\theta)=\int_0^{w_\mathrm{H}}\d w\,
  p_\kappa(w)\,\delta[f_K(w)\vec\theta,w]\;,
\end{equation}
where $p_\kappa(w)$ is the projector
\begin{eqnarray}
  p_\kappa(w)&=&\frac{3}{2}\,\Omega_0\,\left(\frac{H_0}{c}\right)^2
  \nonumber\\&\times&
  \int_w^{w_\mathrm{H}}\frac{\d w'}{a(w')}\,W_\mathrm{Q}(w')\,
  \frac{f_K(w)\,f_K(w-w')}{f_K(w')}\;.
\label{eq:4}
\end{eqnarray}
The ratio between the angular diameter distances $f_K$ is the usual
effective lensing distance, $W_\mathrm{Q}(w)$ is the normalised distance
distribution of the sources, in our case the quasars, and $a(w)$ is
the cosmological scale factor.

The correlation function $w_{\kappa\delta}$ appearing on the
right-hand side of Eq.~(\ref{wqg}) can now be related to the power
spectrum of the density contrast $\delta$,
\begin{eqnarray}
  w_{\kappa\delta}(\theta)&=&\langle
    \kappa(\vec\phi)\,\bar\delta(\vec\phi+\vec\theta)
  \rangle\nonumber\\&=&
  \int\d w\,p_\kappa(w)\int\d w'\,p_\delta(w')\,
  \nonumber\\&\times&\langle
    \delta[f_K(w)\vec\theta,w]\,\delta[f_K(w')(\vec\phi+\vec\theta),w']
  \rangle\;,
\label{eq:6}
\end{eqnarray}
where the integrations over $w$ and $w'$ range from $0$ to
$w_\mathrm{H}$.

We can now use Limber's equation for the statistics of projected
homogeneous and isotropic random fields. Inserting the Fourier
transform of the density contrast, and introducing its power spectrum
$P_\delta(k)$, we find
\begin{eqnarray}
  w_{\kappa\delta}(\theta)&=&\int\d w\,
  \frac{p_\kappa(w)\,p_\delta(w)}{f_K^2(w)}\,\nonumber\\
  &\times&
  \int\frac{s\d s}{2\pi}\,P_\delta\left(\frac{s}{f_K(w)},w\right)\,
    \mathrm{J}_0(s\theta)\;.
\label{expression_w}
\end{eqnarray}

\section{The three-point correlation function}
\label{3-point}

\subsection{Formalism}

Let us now extend the formalism introduced in the previous section to 
 define higher-order statistical quantities. There
are two possibilities for defining a three-point correlator between
quasars and galaxies, either through correlations between single
quasars and galaxy pairs, $z_\mathrm{QGG}$, or between quasar pairs
and single galaxies, $z_\mathrm{QQG}$.

By definition, three-point correlations vanish for Gaussian random
fields. They are created in the course of the non-linear evolution of
the underlying density field, hence they are expected to appear
preferentially on small angular
scales. Since the mean separation between quasars is in general much
larger than between galaxies, correlations between quasars and
galaxy pairs should be much easier to measure. We will therefore focus
on the triple correlator $z_\mathrm{QGG}$ only.

By definition, and using the formalism introduced in the previous
section, we have
\begin{equation}
  z_\mathrm{QGG}(\vec\theta_1,\vec\theta_2)=\langle
    \delta\mu^{\alpha-1}(\vec\phi)\,
    \bar\delta_\mathrm{G}(\vec\phi+\vec\theta_1)\,
    \bar\delta_\mathrm{G}(\vec\phi+\vec\theta_2)\rangle\;,
\label{expression_triple}
\end{equation}
where the average extends over all positions $\phi$.

Assuming as before a linear biasing relation between the galaxies and
the density fluctuations, and expanding the
Eq.~(\ref{expression_triple}) to first order in $\kappa$ and $\gamma$
leads to
\begin{equation}
  z_\mathrm{QGG}(\theta_1,\theta_2)=2\,{\bar b}^2\,(\alpha-1)\,
  z(\vec\theta_1,\vec\theta_2)\;,
\end{equation}
where $z(\vec\theta_1,\vec\theta_2)$ is the three-point correlation
function
\begin{eqnarray}
  z(\vec\theta_1,\vec\theta_2)&=&\langle
    \kappa(\vec\phi)\,\bar\delta(\vec\phi+\vec\theta_1)\,
    \bar\delta(\vec\phi+\vec\theta_2)\rangle\nonumber\\&=&
  \int\d w_1\,p_\kappa(w_1)\int\d w_2\,p_\delta(w_2)
  \int\d w_3\,p_\delta(w_3)\nonumber\\&\times& 
  \left\langle
    \delta[f_K(w_1)\vec\phi,w_1]\,
    \delta[f_K(w_2)(\vec\phi+\vec\theta_1),w_2]
  \right.\nonumber\\&\times&\left.
    \delta[f_K(w_3)(\vec\phi+\vec\theta_2),w_3]
  \right\rangle\;.
\label{expression_z}
\end{eqnarray}
Again, the integrations over $w_{1,2,3}$ extend from $0$ to
$w_\mathrm{H}$. Stochastic or non-linear biasing schemes will be
discussed in Sect.~\ref{mesuring_omega}.

The three-point correlation function
$z_\mathrm{QGG}(\vec\theta_1,\vec\theta_2)$ is related to the excess
probability with respect to a random distribution 
  for finding triangle configurations, defined
by the two angular separation vectors $\vec\theta_1$ and
$\vec\theta_2$, formed by one quasar and two galaxies. As shown by
M\'enard \& Bartelmann (2002), the lensing-induced quasar-galaxy
cross-correlation function $w_\mathrm{QG}$ has a small amplitude,
typically on the order of a few per cent at angular scales of a few
arc minutes. Therefore, in order to achieve a significant signal-to-noise ratio
 for a  higher-order correlation function such as
$z_\mathrm{QGG}$, a very large number of objects will be necessary. 
Moreover, focusing only on particular triangle configurations (for instance
equilateral, isosceles or any other) will
dramatically restrain the number of possible measurements in a given
survey. For our purpose of measuring a skewness parameter (see
Sect. \ref{mesuring_omega}), the detailed angular dependence of a
given triangle configuration is not of immediate relevance. Thus,
it is observationally preferable to focus on an angular average of
$z_\mathrm{QGG}(\vec\theta_1,\vec\theta_2)$ over all suitable triangle
configurations inside a given aperture, thus allowing a measurement
around each quasar. This point will be detailed in
the next Section, where we will also show how the averaged three-point
correlation function can be observed.

\subsection{Observational signature}
\label{observational_signature}

The statistical meaning of $z_\mathrm{QGG}(\vec\theta_1,\vec\theta_2)$
is illustrated by the expression for the \emph{rms} fluctuations in
the counts of galaxies around a given quasar. Following Fry \& Peebles
(1980), and considering projected (rather than three-dimensional)
correlation functions, the variance of the galaxy counts
in cells of solid angle $\mathcal{S}$ at fixed distance from the
quasars can be written
\begin{eqnarray}
  \langle N^2\rangle-\langle N\rangle^2&=&
  \langle N\rangle+n_\mathrm{G}^2\,
  \int_\mathcal{S}\d^2\vec\theta\,\d^2\vec\theta'\,\left[
  w_\mathrm{GG}(\vec\theta-\vec\theta')
  \right.\nonumber\\&+&\left.
  z_\mathrm{QGG}(\vec\theta,\vec\theta')-
  w_\mathrm{QG}(\vec\theta)w_\mathrm{QG}(\vec\theta')\right]\;,
\end{eqnarray}
where the angular brackets denote averages over ensembles of cells of
size $\mathcal{S}$. Thus, we have
$\langle N\rangle=n_\mathrm{G}\mathcal{S}+
 n_\mathrm{G}\int_\mathcal{S}\d^2\vec\theta\,w_\mathrm{QG}(\vec\theta)$,
where $n_\mathrm{G}$ is the galaxy number density on the sky.

If the cells were \emph{randomly} placed rather than at a fixed
position relative to a quasar, the variance of the galaxy counts would
be
\begin{equation}
  \langle N^2\rangle_\mathrm{r}-\langle N\rangle_\mathrm{r}^2=
  \langle N\rangle_\mathrm{r}+n_\mathrm{G}^2\,
  \int_\mathcal{S}\d^2\vec\theta\,\d^2\vec\theta'\,
  w_\mathrm{GG}(\vec\theta'-\vec\theta)\;,
\end{equation}
with $\langle N\rangle_\mathrm{r}=n_\mathrm{G}\mathcal{S}$.

The normalised extra variance $\Delta(\theta)$ of galaxy counts in
cells of area $\mathcal{S}$ near quasars is directly observable and
can be expressed in terms of the integrated correlation functions
 introduced above,
\begin{eqnarray}
  \Delta(\theta)&\equiv&\frac{\langle N^2\rangle-\langle N\rangle^2}
    {\langle N\rangle_\mathrm{r}^2}-
  \frac{\langle N^2\rangle_\mathrm{r}-\langle N\rangle_\mathrm{r}^2}
    {\langle N\rangle^2_\mathrm{r}}\,\nonumber\\&=&
  \bar{z}_\mathrm{QGG}(\theta)-\bar{w}^2_\mathrm{QG}(\theta)+
  \frac{\bar{w}_\mathrm{QG}(\theta)}{\langle N\rangle_\mathrm{r}}\;,
\label{delta}
\end{eqnarray}
where $\bar{w}_\mathrm{QG}$ and $\bar{z}_\mathrm{QGG}$ are the
cell-averaged correlation functions
\begin{equation}
  \bar{w}_\mathrm{QG}(\theta)=
  \frac{\int\d^2\vec\theta'\,U_\theta(\vec\theta')\,
    w_\mathrm{QG}(\vec\theta')}
  {\int\d^2\vec\theta'\,U_\theta(\vec\theta')}
\label{integ_w}
\end{equation}
and
\begin{equation}
  \bar{z}_\mathrm{QGG}(\theta)=
  \frac{\int\d^2\vec\theta_1\,U_\theta(\vec\theta_1)\,
        \int\d^2\vec\theta_2\,U_\theta(\vec\theta_2)\,
        z_\mathrm{QGG}(\vec\theta_1,\vec\theta_2)}
  {\int\d^2\vec\theta_1\,U_\theta(\vec\theta_1)\,
   \int\d^2\vec\theta_2\,U_\theta(\vec\theta_2)}\;.
\label{integ_z}
\end{equation}
Here, the function $U_\theta(\theta')$ is a top-hat filter of radius
$\theta$. Other filters can be used if one wishes to be more sensitive
to specific wavelength ranges of the dark-matter power spectrum (see
Schneider et al.~1998) but we will not investigate this point in the
present paper.

In summary, the relevant third-order quantity for our purpose is the cell
average $\bar{z}_\mathrm{QGG}(\theta)$ of the triple correlator
$z_\mathrm{QGG}(\vec\theta_1,\vec\theta_2)$ between quasars and galaxy
pairs. We will now estimate its properties and
then show how second and third-order quasar-galaxy correlations can be
combined to measure $\Omega_0$ and several properties of the galaxy
bias.

\section{Evaluation of the excess galaxy variance $\Delta$}
\label{evaluation}

\subsection{Statistics of the density field}

Under the common assumption that the initial density fluctuations were
Gaussian and that cosmic structure grew by gravitational instability,
the three-point correlation function is intrinsically a second-order
quantity, and should be detectable only where non-linearities arise in
the density field.

We expand the density field to second order as
\begin{equation}
  \delta(\vec{x})=\delta^{(1)}(\vec{x})+\delta^{(2)}(\vec{x})\;,
\end{equation}
where $\delta^{(2)}$ is of order $(\delta^{(1)})^2$ and represents
departures from Gaussian behaviour. Note that since two of the three
$\delta$ factors in Eq.(\ref{expression_z}) are derived from the
fluctuations in the galaxy number density, this expansion is only
valid if galaxies are linearly biased relative to the dark-matter
fluctuations.
For now we will assume this linearity and show in
Sect.~\ref{mesuring_omega} how cosmic magnification can be used to
measure the matter density parameter. 
(This property of linearity has been measured on scales between 5 and
30$h^{-1}$Mpc with the 2dF Galaxy Redshift Survey (Verde et al. 2002)).
In next section we will also show that
quasar-galaxy correlations can also test this property and
probe the angular range where this simple relation between dark matter
and galaxies breaks down.

Assuming linear biasing and expanding $\delta$ to second order, we
obtain for $\langle\delta_1\delta_2\delta_3\rangle$
\begin{eqnarray}
  \langle\delta_1\delta_2\delta_3\rangle&\simeq&
  \langle\delta_1^{(1)}\delta_2^{(1)}\delta_3^{(1)}\rangle+
  \langle\delta_1^{(1)}\delta_2^{(1)}\delta_3^{(2)}\rangle
  \nonumber\\&+&
  \mbox{cyclic terms (231,312)}\;.
\label{developement}
\end{eqnarray}
The first term in Eq.~(\ref{developement}) vanishes because the
density fluctuation field is Gaussian to first order, hence the third
moment of $\delta^{(1)}$ is zero. Thus, the leading term in
Eq.~(\ref{developement}) is of the order of
$\langle\delta_1^{(1)}\delta_2^{(1)}\delta_3^{(2)}\rangle$.

In second-order perturbation theory, the Fourier decomposition of the
density-fluctuation field is given by
\begin{eqnarray}
  \delta^{(2)}(\vec{k})&=&
  \int\d^3\vec{k_1}\,d^3\vec{k_2}\,\delta^{(1)}(k_1)\delta^{(1)}(k_2)
  \nonumber\\&\times&
  \delta_\mathrm{D}(\vec{k_1}+\vec{k_2}-\vec{k})\,
  F(\vec{k_1},\vec{k_2})\;,
\end{eqnarray}
with
\begin{equation}
  F(\vec{k_1},\vec{k_2})=\frac{5}{7}+
    \frac{1}{2}\frac{\vec{k_1}\cdot\vec{k_2}}{k_1\,k_2}\,
    \left(\frac{k_1}{k_2}+\frac{k_2}{k_1}\right)+
    \frac{2}{7}\frac{(\vec{k_1}\cdot\vec{k_2})^2}{k_1^2k_2^2}\;.
\label{deltadef}
\end{equation}

Note that we have expanded the lensing-induced magnification to first
order in Eq.~(\ref{expression_z}). The next-order term has a
contribution proportional to $\kappa^2$ which, when introduced in
Eq.~(\ref{developement}), is of order $(\delta^{(1)})^4$ and thus
formally of the same order as the other terms in
Eq.~(\ref{developement}). However, this additional term differs by
the weight factor $f_K(w-w')/f_K(w')$ from the other terms since it
contains the lensing efficiency.  As a result, this additional factor
will be one order of magnitude smaller, even though it is of the same
order in the perturbation series for $\delta$ (see Van Waerbeke et
al.~2001 for more detail). We can therefore neglect its contribution.

\subsection{Bispectrum and Non-Linear evolution}

\begin{figure*}[ht]
\begin{center}
  \includegraphics[width=.45\hsize]{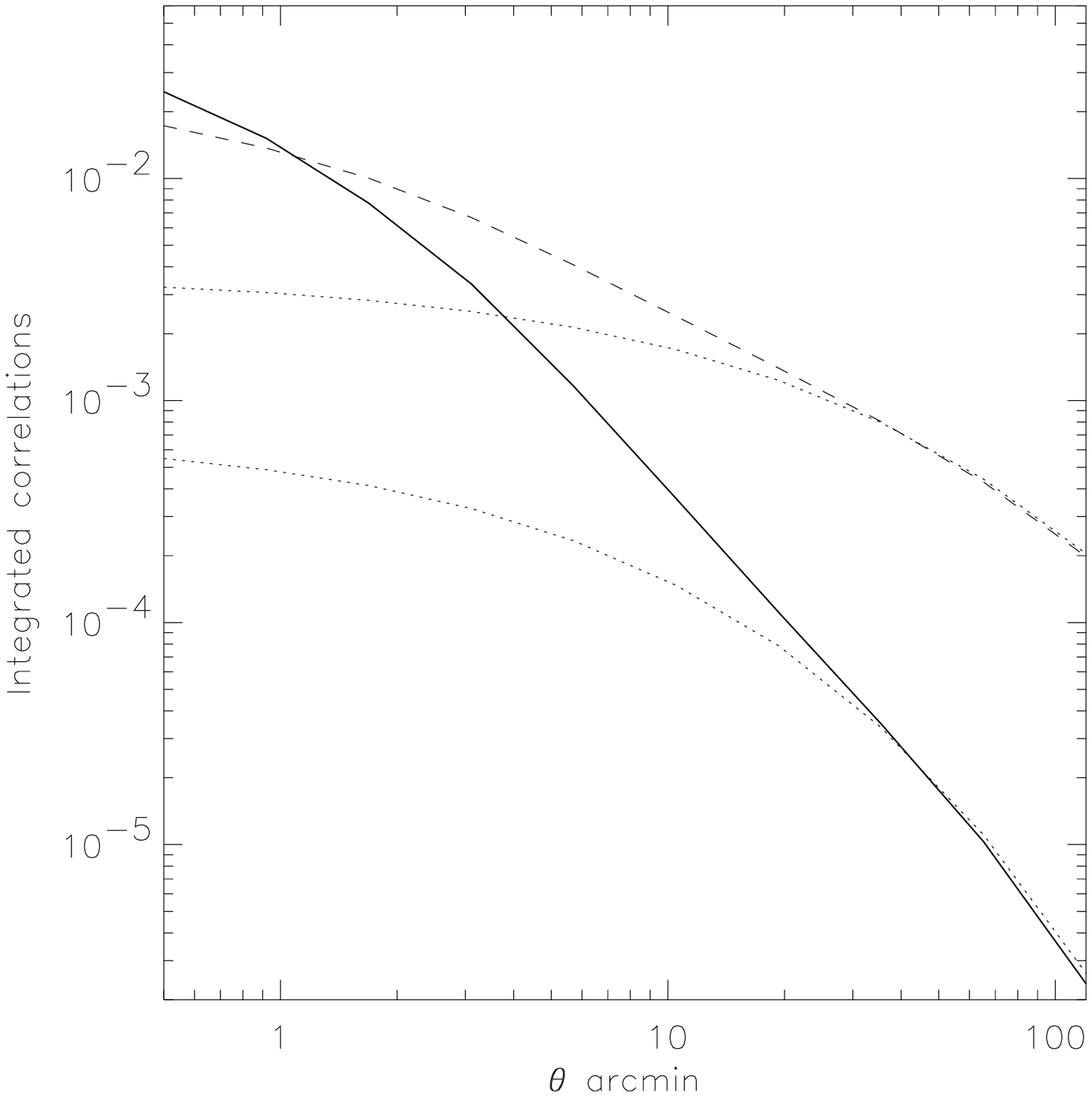}\hfill
  \includegraphics[width=.45\hsize]{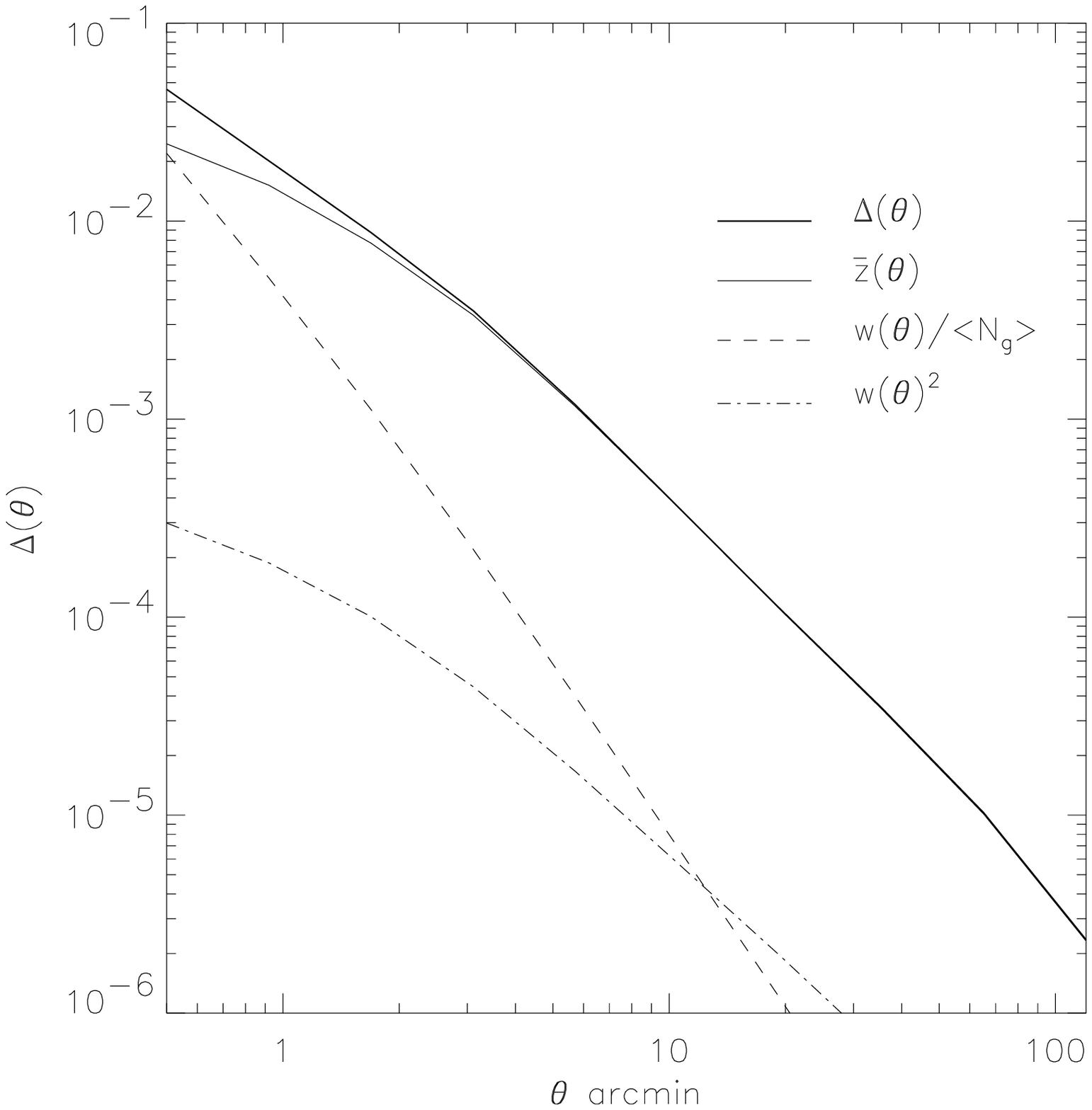}
\caption{\emph{Left panel}: The two- and three-point quasar-galaxy
  correlation functions, averaged within disks of radius $\theta$, are
  plotted as a function of angular scale for a flat Universe with
  $\Omega_0=0.3$.  The dashed curve shows the two-point correlation
  function $\bar{w}(\theta)$ and the solid line the three-point
  function $\bar{z}(\theta)$. In dotted lines are plotted the
  corresponding correlation functions for a linear evolution of the
  perturbation. Near one arc minute, the triple correlator $\bar{z}$
  reaches the $\sim2\%$ level, assuming a linear bias of unity. At
  larger angular scales, the amplitude drops steeply.
\emph{Right panel}: We plot the expected extra-scatter of galaxies
  around quasars for a flat Universe with $\Omega_0=0.3$ (thick line)
  and considering a galaxy density of 1 per arcmin$^2$. We also show
  the different contributions of Eq. \ref{delta}.}
\label{z}
\end{center}
\end{figure*}

The ensemble average in Eq.~(\ref{developement}) is related to the
bispectrum in Fourier space. By definition,
\begin{eqnarray}
  \langle\delta(\vec{k_1})\delta(\vec{k_2})\delta(\vec{k_3})\rangle
  &=&(2\pi)^3\,B(\vec{k_1},\vec{k_2},\vec{k_3})\nonumber\\
  &\times&
  \delta_\mathrm{D}(\vec{k_1}+\vec{k_2}+\vec{k_3})\;.
\label{def_bispectrum}
\end{eqnarray}
Inserting Eq.~(\ref{deltadef}) into Eq.~(\ref{def_bispectrum}) leads
to an expression of the bispectrum in terms of the second-order kernel
$F(\vec k_1,\vec k_2)$ and the dark-matter power spectrum
\begin{eqnarray}
  B(\vec k_1,\vec k_2,\vec k_3)&=&
  2\,F(\vec k_1,\vec k_2)\,P(k_1)P(k_2)\nonumber\\&+&
  2\,F(\vec k_2,\vec k_3)\,P(k_2)P(k_3)\nonumber\\&+&
  2\,F(\vec k_1,\vec k_3)\,P(k_1)P(k_3)\;.
\label{bispectredef}
\end{eqnarray}
Note that Eq.~(\ref{def_bispectrum}) implies that the bispectrum is
nonzero only if the wave vectors $(\vec k_1,\vec k_2,\vec k_3)$ form
closed triangles.

For describing the bispectrum on all angular scales, we use the
fitting formula for the non-linear evolution of the bispectrum derived
from numerical CDM models by Scoccimarro \& Couchman (2001), extending
earlier work assuming scale-free initial conditions. The kernel
$F(\vec k_1,\vec k_2)$ in Eq.~(\ref{bispectredef}) is then simply
replaced by an \emph{effective} kernel $F^\mathrm{eff}(\vec k_1,\vec
k_2)$, reading
\begin{eqnarray}
  F^\mathrm{eff}(\vec k_1,\vec k_2)&=&
  \frac{5}{7}\,a(n,k_1)a(n,k_2)\nonumber\\&+&
  \frac{1}{2}\,\frac{\vec k_1\cdot\vec k_2}{k_1 k_2}\,
  \left(\frac{k_1}{k_2}+\frac{k_2}{k_1}\right)\,b(n,k_1)b(n,k_2)
    \nonumber\\&+&
  \frac{2}{7}\left(\frac{\vec k_1\cdot\vec k_2}{k_1 k_2}\right)^2\,
    c(n,k_1)c(n,k_2)\;.
\label{F_expression}
\end{eqnarray}
The coefficients $a$, $b$ and $c$ are given by
\begin{eqnarray}
  a(n,k)&=&
  \frac{1+\sigma_8^{-0.2}(z)\left[0.7\,Q_3(n)\right]^{1/2}(q/4)^{n+3.5}}
       {1+(q/4)^{n+3.5}}\nonumber\\
  b(n,k)&=&
  \frac{1+0.4\,(n+3)\,q^{n+3}}{1+q^{n+3.5}}\nonumber\\
  c(n,k)&=&
  \frac{1+4.5/\left[1.5+(n+3)^4\right](2q)^{n+3}}{1+(2q)^{n+3.5}}\;,
\end{eqnarray}
where $n=\d\log P_\mathrm{lin}(k)/\d\log k$ is the local slope of the
dark-matter power spectrum at wave number $k$, $q\equiv
k/k_\mathrm{NL}(z)$, and $k_\mathrm{NL}$ is the nonlinear wave number
defined by $4\pi k_\mathrm{NL}^3\,P_\mathrm{lin}(k_\mathrm{NL})=1$. To
be specific, $P_\mathrm{lin}(k)$ is the linear power spectrum at the
required redshift. Finally, the function $Q_3(n)$ is given by
\begin{equation}
  Q_3(n)=\frac{(4-2^n)}{(1+2^{n+1})}\;,
\label{Q3def}
\end{equation}
which is the so-called saturation value obtained in ``hyper-extended''
perturbation theory (HEPT; Scoccimarro \& Frieman 1999). These
expressions imply that at large scales, where the coefficients $a$,
$b$ and $c$ aproach unity, the tree-level perturbation theory is
recovered. On the other hand, at small scales, where $a^2\to
(7/10)Q_3(n)\sigma_8^{-0.4}$, $b\to0$ and $c\to0$, the bispectrum
becomes hierarchical with an amplitude which approximately reproduces
HEPT for $\sigma_8\approx1$. For more detail, see Scoccimarro \&
Couchman (2001).

\subsection{Evaluation of $z_\mathrm{QGG}(\theta)$}

The formalism is now in place for computing the triple
correlator $z(\theta_1,\theta_2)$. We start from
Eq.~(\ref{expression_z}) and replace the density contrast $\delta$ by
its Fourier transform. Next, we employ the approximation underlying
Limber's equation, which asserts that the coherence length of the
density fluctuation field is much smaller than the scales on which both
projectors $p_\kappa$ and $p_\delta$ vary appreciably. Finally, we
insert the expression for the bispectrum described in the previous
section and find
\begin{eqnarray}
  z(\theta_1,\theta_2)&=&\int\d w\,p_\kappa(w)\,p_\delta^2(w)
  \int\frac{\d^2\vec k_1}{(2\pi)^2}\,
  \mathrm{e}^{i\vec{k_1}\cdot\vec{\theta}_1\,f_K(w)}
  \nonumber\\&\times&
  \int\frac{\d^2\vec k_2}{(2\pi)^2}\,B_\delta(k_1,k_2,-\vec k_1-\vec k_2,w)\,
  \mathrm{e}^{i\vec{k_2}\cdot\vec{\theta}_2\,f_K(w)}\;.\nonumber
\end{eqnarray}

We saw in Sect.~3 that the suitable quantity for observations is not
directly $z(\vec\theta_1,\vec\theta_2)$, but the corresponding
cell-averaged triple correlator $\bar{z}(\theta)$ defined by
Eq.~(\ref{integ_z}). The angular integration yields
\begin{eqnarray}
  \bar{z}(\theta)&=&4\,\int\d w\,p_\kappa(w)\,p_\delta^2(w)\,
  \int\frac{\d^2 \vec k_1}{(2\pi)^2}\,
  \frac{\mathrm{J}_1[k_1\theta\,f_K(w)]}{k_1\theta\,f_K(w)}
  \\&\times&
  \int\frac{\d^2 \vec k_2}{(2\pi)^2}\,B_\delta(k_1,k_2,-\vec k_1-\vec k_2,w)\,
  \frac{\mathrm{J}_1[k_2\theta\,f_K(w)]}{k_2\theta\,f_K(w)}\nonumber
\label{z_expression}
\end{eqnarray}

The observational signature of cosmic magnification is expressed in
terms of the normalised excess scatter $\Delta(\theta)$ of galaxies
around quasars; see Eq.~(\ref{delta}). This quantity involves both the
two- and three-point cell-averaged correlation functions
$\bar{w}_\mathrm{QG}(\theta)$ and $\bar{z}_\mathrm{QGG}(\theta)$. For
numerically evaluating these expressions, we use the description of
the non-linear 
  evolution of the CDM power spectrum provided by
 Peacock \& Dodds (1996) fitting formula.  We normalise the power
spectrum such that the local abundance of galaxy clusters is
reproduced, $\sigma_8=0.52\,\Omega_0^{-0.52+0.13\,\Omega_0}$, as
determined by Eke et al.~(1996).

We assume for simplicity that all quasars are at the same redshift
$z_\mathrm{s}=1.5$. More realistic quasar redshift distributions do not
significantly change the following results as long as the foreground
galaxies are at comparatively low redshift. We approximate the
redshift distribution of the galaxies by
\begin{equation}
  p_\delta(z)\,\d z=\frac{\beta\,z^2}{z_0^3\,\Gamma(3/\beta)}\,
  \exp\left[-\left(\frac{z}{z_0}\right)^\beta\right]\,\d z\;,
\end{equation}
with $\beta=1.5$ and $z_0=0.3$.

The slope of the quasar number counts is fairly well constrained 
  by the most recent quasar catalogues.  We use the value $\alpha=2$
suggested by the first SDSS quasar catalogue (Schneider et al. 2001)
for quasars brighter than $19$th magnitude. Finally, we assume $\bar b=1$
for simplicity.

The left panel of Fig.~\ref{z} shows $\bar{z}_\mathrm{QGG}(\theta)$
(solid line) and $\bar{w}_\mathrm{QG}(\theta)$ (dashed line) as a
function of angular scale. These quantities were computed for a
low-density, spatially flat universe ($\Omega_0=0.3$,
$\Omega_\Lambda=0.7)$.  The dotted lines show the expected amplitude
of $\bar{w}_\mathrm{QG}(\theta)$ if only linear growth of density
perturbations is taken into account, and the amplitude of
$\bar{z}_\mathrm{QGG}(\theta)$ for quasi-linear evolution. The
difference between the two regimes changes the amplitude of
$\bar{z}_\mathrm{QGG}$ by approximately two orders of magnitude on
small angular scales. On large scales, the amplitude of
$\bar{z}_\mathrm{QGG}(\theta)$ decreases quickly with $\theta$ since
the density field tends to gaussianity as the smoothing scale
increases, and thus the triple correlator vanishes.

Interestingly, the amplitude of $\bar{z}_\mathrm{QGG}$ is of the order
of one per cent on arcminute scales. As for the two-point
quasar-galaxy correlation function, the amplitude and the shape of
$\bar{z}_\mathrm{QGG}(\theta)$ are very sensitive to cosmological
parameters. 

In the right panel of Fig.~\ref{z}, we plot the measurable quantity
$\Delta(\theta)$ which represents the normalised excess scatter of
galaxies around quasars; cf.~Eq.~(\ref{delta}). Again, we have used
the $\Lambda$CDM cosmological model, and we assume a galaxy number
density of $1\,\mathrm{arcmin}^{-2}$.

Evidently, $\bar{z}_\mathrm{QGG}(\theta)$ is the main contribution to
$\Delta(\theta)$ on intermediate and large angular scales.  Below a
few arcminutes, the term $\bar{w}_\mathrm{QG}(\theta)/\langle
N\rangle_\mathrm{r}$ becomes non-neglegible. This contribution is due
to the shot noise of the galaxies, thus this term can be lowered when
using galaxies with a higher number density.

\section{Determining $\Omega_0$ and testing the linearity of the bias}
\label{mesuring_omega}

The second- and third-order statistics can be used jointly 
 so that several parameter
dependencies can cancel. The underlying physical concept is that
second-order statistics quantify the Gaussian characteristics of a
random process, while third-order statistics are 
non-Gaussian contributions. When used together, one can in principle 
  measure their relative strength, thus isolating those
parameters which are most responsible for deviations from Gaussianity.

\begin{figure}[ht]
\begin{center}
  \includegraphics[width=\hsize]{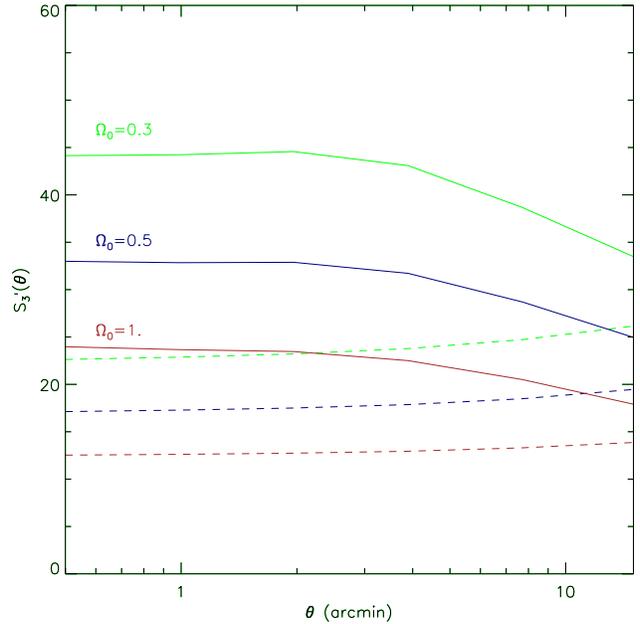}
\caption{The skewness $S'_3(\theta)$ is shown as a function of angle
  $\theta$ for three different cosmological models. The solid line is
  the prediction based on the non-linear description of the power
  spectrum and the bispectrum. The dashed line shows the
  perturbation-theory calculation.} 
\label{s3}
\end{center}
\end{figure}

The reduced skewness (i.e.~the ratio of the third and second 
moments of a distribution) is a useful practical estimate of non-Gaussian
  features in galaxy catalogs. However, in the case of cosmic magnification,
it is not possible to define the skewness in the same way as for
cosmic shear with the convergence field (Bernardeau et al. 1997),
since we are not considering the autocorrelation 
properties of a single field, but the cross-correlations between two
different fields, namely the distributions of foreground galaxies and
of the lensing convergence $\kappa$.
Moreover, the angular-averaged three-point correlation function
$\bar{z}_\mathrm{QGG}$ discussed in the previous section 
  is not symmetric with respect to permutations
between quasar and galaxy positions. The three-point cross-correlation
of quasars with galaxy pairs involves the quasar-galaxy
cross-correlation as well as the galaxy-galaxy auto-correlation. Since
the latter does not contribute to the two-point quasar-galaxy
correlation function, we cannot apply the usual mathematical
definition of skewness which applies to a unique
distribution. Instead, we define a pseudo-skewness for our purposes by
the ratio
\begin{equation}
  S'_3(\theta)=\frac{\bar{z}_\mathrm{QGG}(\theta)}
                    {\bar{w}_\mathrm{QG}^2(\theta)}\;.
\end{equation}
Much like the skewness of cosmic shear, 
  $S'_3(\theta)$ is insensitive in the linear regime 
to the normalisation of the  power spectrum $\sigma_8$, 
 and still remains weakly dependent on $\sigma_8$ even in the
non-linear regime. Moreover, if the galaxy bias can be considered
linear on certain angular scales, we have
\begin{equation}
  S'_3(\theta)= A(\theta)\,\Omega_0^{-n}
\end{equation}
on such scales, 
where $n$ is close to one half and the amplitude
$A(\theta)$ can be computed from 
Eqs.~(\ref{expression_w}), (\ref{expression_triple}) and
(\ref{z_expression}). Interestingly, $S'_3(\theta)$ does not depend on
the linear galaxy bias factor. We plot the pseudo-skewness in Fig.~\ref{s3}
for different flat and cluster-normalised CDM model universes with
density parameters $\Omega_0=0.3$, $0.5$ and $1$. The solid line shows
the result obtained with the non-linear prescription of the
bispectrum, and the dashed line shows the result of using quasi-linear
theory.

\begin{figure*}[ht]
\label{s3_sigma8_Gamma}
\begin{center}
  \includegraphics[width=.45\hsize]{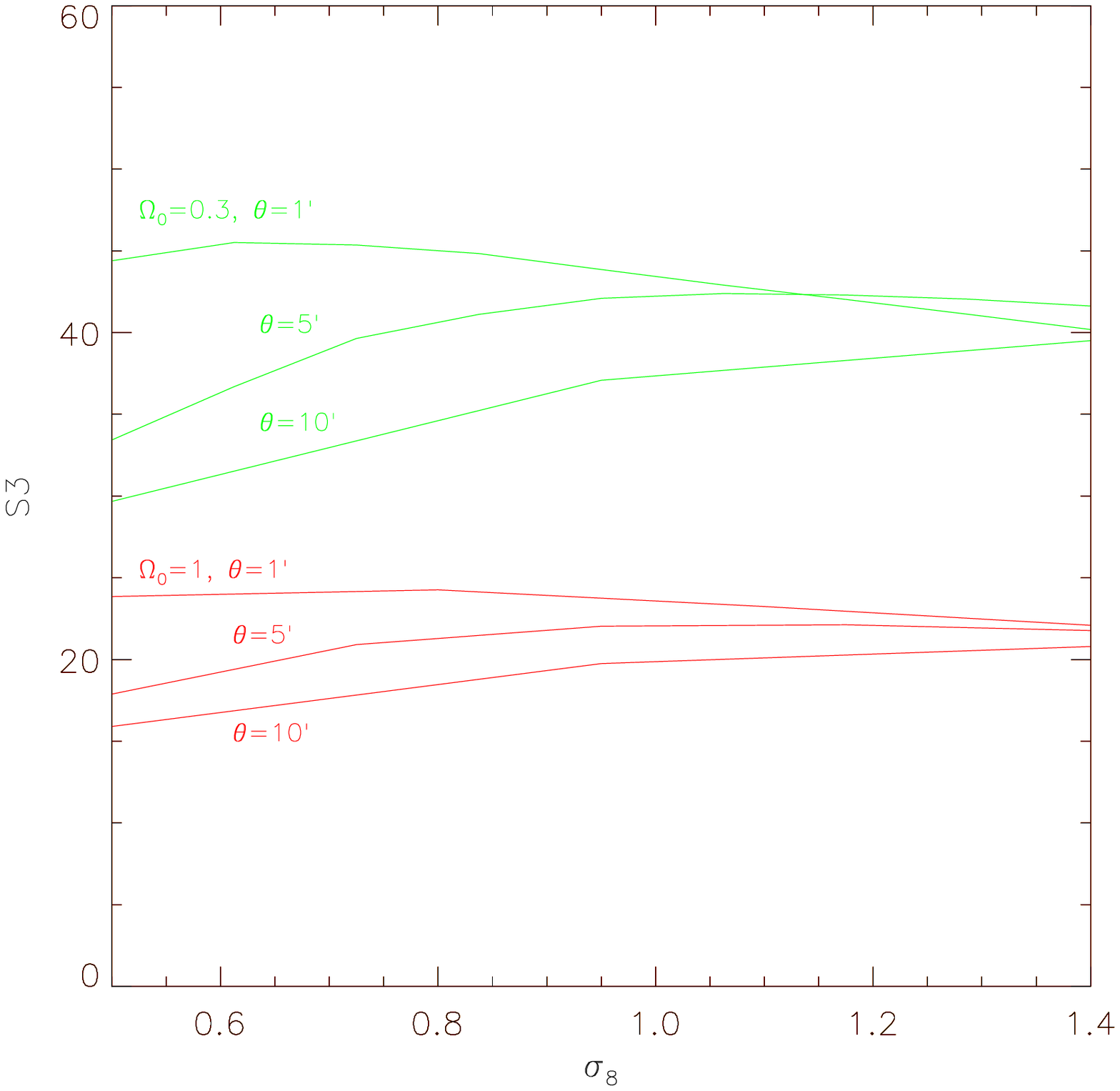}
  \includegraphics[width=.45\hsize]{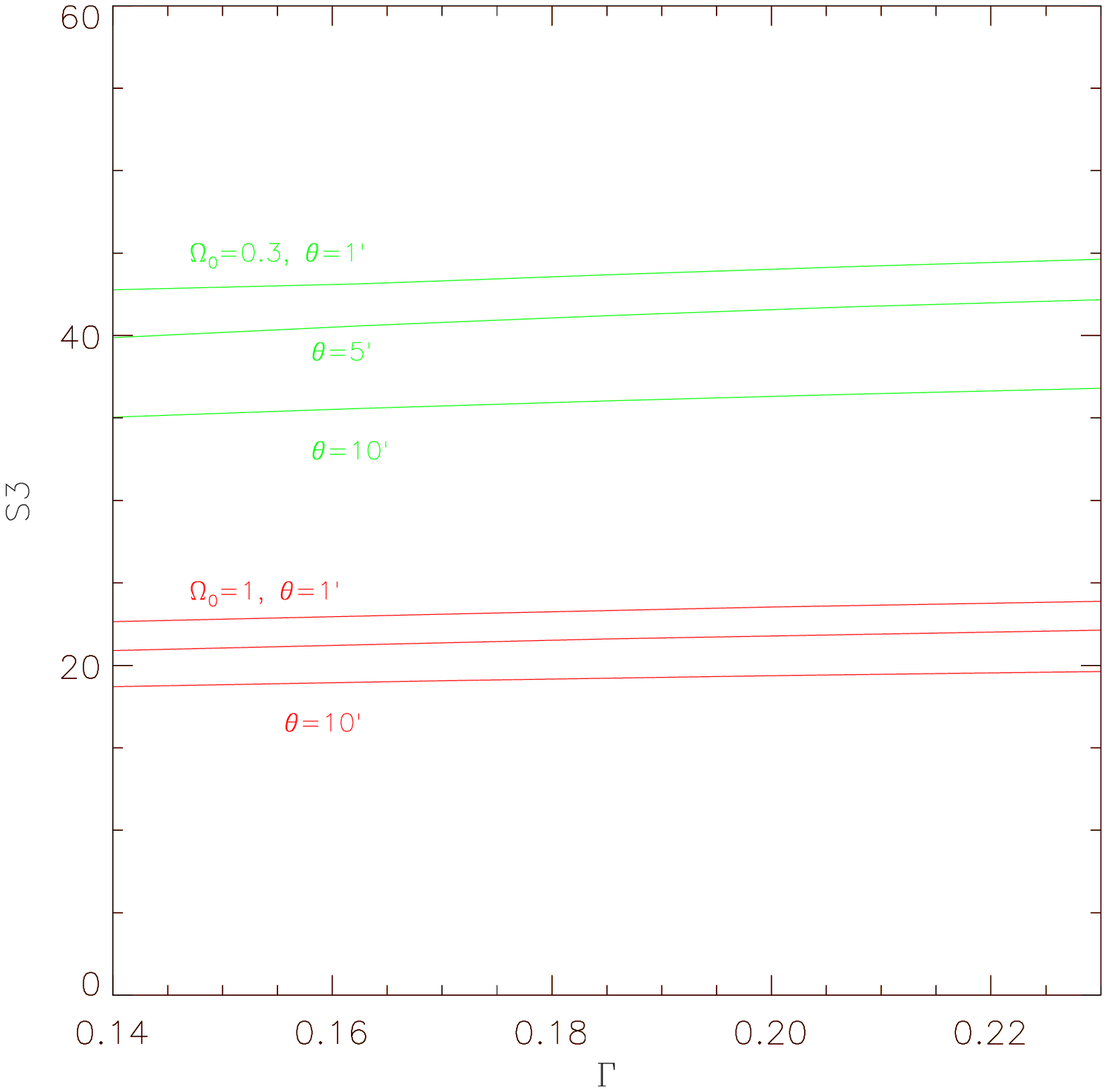}
\caption{The dependence of the skewness $S_3'$ on $\sigma_8$ and
  $\Gamma$ is shown for two different cosmologies ($\Omega_0=1$ and
  $0.3$) and three different
  angular scales, $1'$, $5'$ and $10'$. 
A value of $\Gamma=0.21$ is used when $\sigma_8$ varies and we fix
$\sigma_8=0.9$ when $\Gamma$ varies.
The dependence of $S'_3$ on $\sigma_8$ goes down to $10\%$ at small scales
and is at the one percent level when $\Gamma$ changes.}
\end{center}
\end{figure*}

The non-linear growth of density perturbations introduces some
dependency of the pseudo-skewness $S'_3(\theta)$ on the normalisation
$\sigma_8$ and the shape parameter $\Gamma$ of the dark-matter power
spectrum because $\bar{w}_\mathrm{QG}$ and $\bar{z}_\mathrm{QGG}$
depend on the shape of the power spectrum in different way;
cf.~Eqs.~(\ref{expression_w}) and (\ref{expression_z}). We plot these
dependences in Fig.~\ref{s3_sigma8_Gamma}.

The figure shows that $S'_3(\theta)$ depends relatively weakly on
$\sigma_8$ and is almost insensitive to $\Gamma$. 
Varying $\sigma_8$ from $0.5$ to $1.4$ changes the amplitude of 
$S'_3(\theta)$ by $30\%$ near $10$ and $5$
arcminutes and this change in the amplitude reduces to $10\%$ near one
arc minute. Therefore we see that the scales of a few arcminutes will be
favored for estimating $\Omega_0$.
The effect of changing $\Gamma$ from $0.14$ to $0.23$
is even weaker and does not depend on the angular scale as well as the
value of $\Omega_0$. Considering the
above mentionned range the $S'_3$ dependence on $\Gamma$ is at the one 
percent level.
Moreover, if additional measurements allow to reduce the possible ranges for
$\sigma_8$ and $\Gamma$, the dependence of $S'_3$ on these parameters
further reduces.
Thus, we can consider the pseudo-skewness
$S'_3(\theta)$ to be weakly sensitive to normalisation of the 
dark-matter power spectrum, and almost 
insensitive to its shape parameter. 
Likewise, the cosmological constant
$\Omega_\Lambda$ has a negligible effect.

On large angular scales, the galaxy bias is expected to be linear (Verde et
al. 2002), in which case its contribution to $S'_3$ cancels out.
Finally, the only effective parameter we are left with is $\Omega_0$,
which means that the pseudo-skewness $S_3'(\theta)$ can effectively
constrain the matter-density parameter.

The pseudo-skewness $S_3'(\theta)$ can also test the linearity and
stochasticity of the bias parameter which is of interest at smaller
scales. For a given cosmology and assuming a CDM power spectrum,
any departure from the angular
variation of $S_3'(\theta)$ given in Fig.~\ref{s3} can be interpreted
as resulting from
non-linearity and/or stochasticity of the bias. Indeed, our previous
calculations deriving the two- and three- point correlators used only
two hypotheses, namely the linearity of the bias 
and the assumption that lensing effects occur in the weak regime, i.e. 
$\delta\mu=2\,\kappa+\mathcal{O}(\kappa^2,|\gamma^2|)$. 
The latter relation is expected to be accurate on scales larger than a few
arc minutes. Below, medium- and strong-lensing effects become
non-negligible. Moreover, as described in M\'enard et al.~(2002),
these effects can be quantified by expanding the 
Taylor series of the magnification to second order, thus allowing the
investigation of smaller scales.
Therefore, the only remaining explanation for a departure of the
angular variation of $S_3'(\theta)$ is a nonlinearity or
stochasticity of the biasing scheme. 

The pseudo-skewness can thus
probe the angular range and the corresponding physical scale where the
linear relation between dark matter and galaxy fluctuations breaks down.

\section{Expected signal-to-noise ratio}
\label{section_sn}

We now estimate the expected signal-to-noise ratio in measurements of
$\Delta(\theta)$.  The determination of the effective skewness also
requires a measurement of the two-point quasar-galaxy correlation
function. We refer the reader to M\'enard \& Bartelmann (2002) for a
detailed study of the signal-to-noise ratio expected for
$w_\mathrm{QG}(\theta)$.\\
For the measurement of $\bar{z}_\mathrm{QGG}(\theta)$ we are not
aiming at a detailed noise calculation, but rather an approximate
estimation of the main source of error, i.e.~the finite sampling error
caused by the limited number of available quasars. As shown in
Fig.~\ref{s3}, the dominant contribution to the excess scatter
$\Delta(\theta)$ introduced in Eq.~(\ref{delta}) is
$\bar{z}_\mathrm{QGG}(\theta)$, except on small scales. We therefore
use the simplifying assumption that $\Delta\sim\bar{z}_\mathrm{QGG}$
in the following.

Since the excess scatter of galaxies around quasars defined in
Eq.~(\ref{delta}) is a counts-in-cells estimator, its measurement
accuracy will be limited by the finite size of the available sample,
by boundary and edge effects, and by the effects of discrete
sampling. In practice, measurements of $\bar{z}_\mathrm{QGG}(\theta)$
will be restricted to angular scales much smaller than the size of the
survey, thus the errors contributed by boundary effects can be
considered negligibly small compared to the finite sampling of the
galaxies, which causes the main limitation.

In order to estimate this noise, we assume the quasars to be at random
positions in the sky, i.e.~uncorrelated with the galaxy positions. In
fact, physical correlations are excluded given the required large
separation between the two populations, and on the other hand the
cross-correlations between quasars and galaxies induced by lensing are
so weak that the corresponding change in the galaxy distribution is
entirely negligible in the total error budget. We then assume a
realistic distribution of galaxies (see Appendix A) and find the
standard deviation of the normalised scatter of the galaxy counts
around $N_\mathrm{QSO}$ quasar positions to be
\begin{eqnarray}
  \sigma_\theta^2\left(\frac{\langle N^2\rangle-\langle N\rangle^2}
        {\langle N\rangle_\mathrm{r}^2}\right)&=&
	\frac{1}{N_\mathrm{QSO}}\,\Big(\frac{1}{\bar{N}^3}+
	\frac{2+3\,w}{\bar{N}^2}+\frac{4\,\bar w}{\bar N}+\nonumber\\
	&& \frac{8.4 \bar w^2}{\bar N}+2\,\bar w+28.2\,\bar w^3\Big)
\label{sigma}
\end{eqnarray}
where $\bar{N}$ is the average number of galaxies in cells of a radius
$\theta$.  This result is derived in Appendix~A, using the formalism
developed by Szapudi \& Colombi, and considering galaxy properties
measured by SDSS.

\begin{figure}[ht]
  \includegraphics[width=\hsize]{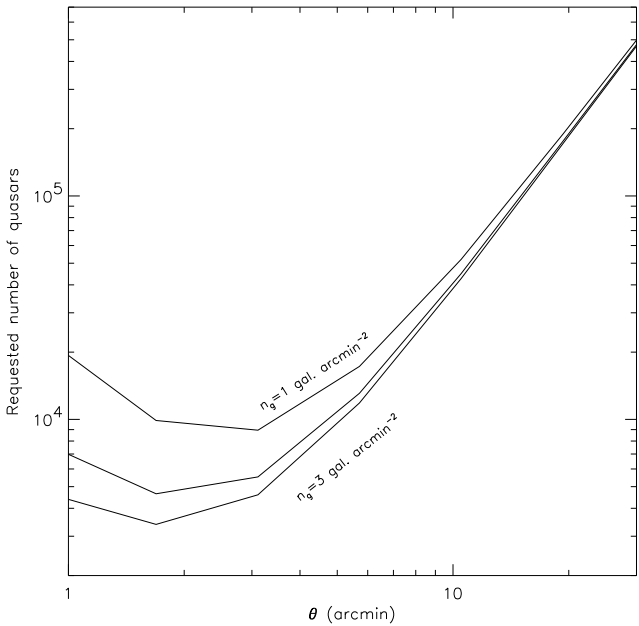}
\caption{Number of quasars required for a 3-$\sigma$ detection of the
  excess scatter $\Delta$ of the galaxy counts near quasars, as a
  function of angular scale $\theta$, and according to our simplified
  noise estimation. Three different values for the
  galaxy number density are assumed, namely $1$, $2$, and
  $3\,\mathrm{arcmin}^{-2}$. The vertical line indicates the mean
  angular separation of $20'$ between distant quasars in the SDSS.}
\label{n_qso}
\end{figure}

The signal we are interested in is an excess scatter in galaxy
counts. Therefore, the noise of the total measurement is twice the
value of $\sigma$ introduced in Eq.~(\ref{sigma}). It is possible to
reduce this noise by measuring the scatter of the galaxy counts at a
larger number of random locations since there are typically many more
available galaxies than quasars, therefore the dominant contribution
comes from the measurements around quasars only. Finally, we can write
the number of required quasars for achieving a detection at an angular
scale $\theta$~:
\begin{equation}
  N_\mathrm{QSO}=\frac{\nu^2}{\left[
    \bar{z}_\mathrm{QGG}(\theta)\,\sigma_\theta\right]^2}\;,
\label{n_qso_simple}
\end{equation}
where $\nu$ is the desired signal-to-noise ratio and $\sigma_\theta$ is the 
error computed from Eq. \ref{sigma}.
We recall that the quasar number given by Eq.~(\ref{n_qso_simple}) refers to
quasars with redshifts higher than those of the galaxy population.

Several weighting schemes can be used to maximise the signal-to-noise
ratio of the detection. M\'enard \& Bartelmann (2002) showed how to
optimally weight the contribution of each quasar with respect to its
magnitude. Weighting the galaxies with respect to their redshift can
also increase the signal-to-noise ratio of the detection. Indeed the
triple correlator $z_\mathrm{QGG}(\vec\theta_1,\vec\theta_2)$ is
related to the excess of triangle configurations in which two galaxies
trace a high density region of the dark matter field,
i.e.~configurations in which the two galaxies are close in angle and
redshift. Projection effects mimicking galaxy pairs will contribute
noise to the final measurement. Therefore, the width of the galaxy
redshift distribution inside a given cell around a quasar can be used
to additionally weight the final measurement. Numerical simulations
will be needed to quantify the change of the final signal-to-noise
ratio, as well as the effects of galaxy clustering and cosmic variance
which were not taken into account in our noise estimation.

As an application, we now investigate the feasibility of measuring
$\Delta(\theta)$ with the data of the \emph{Sloan Digital Sky Survey}
(SDSS~; York et al. 2000). Within this project, the sky has already
been imaged for two years, and the survey will be completed in 2005,
reaching a sky coverage of $\sim10\,000$ square
degrees. Depending on the limiting magnitude of the selected sample,
SDSS can achieve a galaxy density of $\bar{n}_\mathrm{G}\approx
1\,\mathrm{arcmin}^{-2}$ for galaxies observed down to $r'=21$, or
$\bar{n}_\mathrm{G}\approx 3\,\mathrm{arcmin}^{-2}$ down to $r'=22$,
but requiring extensive careful masking of regions with poor seeing
within the survey, and a careful star-galaxy separation (Scranton et
al.~2001).

In Fig.~\ref{n_qso}, we present for different values of the galaxy
number density the number of quasars required in order to achieve a 3-$\sigma$
detection of the excess scatter $\Delta(\theta)$. 
The figure shows
that the measurement becomes more easily reachable at intermediate angular
scales. 
Note that in our estimation we have assumed isolated
quasars and Eq.~\ref{n_qso_simple} ceases to be
valid at angular scales 
reaching the average angular separation of quasars. In the case of
SDSS, this angular scale is close to $20'$ for the sample of
spectroscopic quasars satisfying $z>1$. 
On larger
angular scales, the real errors will increase due to correlated galaxy
counts, and the number of quasars estimated from
Eq.~(\ref{n_qso_simple}) will no longer be reliable. The SDSS will
observe $10^5$ spectroscopic and $5\times10^5$ to $10^6$ photometric
quasars. Thus, we conclude from Fig.~\ref{n_qso} that SDSS will
exceed the number of quasars required for measuring
three-point correlations between quasars and galaxies, thus allowing a
new direct measurement of the matter density $\Omega_0$.

\section{Conclusion}

Via the magnification bias, gravitational lensing by large-scale
structures gives rise to angular cross-correlations between distant
sources and foreground galaxies although the two populations are
physically uncorrelated. Depending on whether the matter along the
lines-of-sight towards these background sources is over- or underdense
with respect to the mean, and depending on the value of the slope
$\alpha$ of the cumulative number counts of the sources, magnification
effects can cause an excess or a deficit of distant sources near
foreground galaxies.

These lensing-induced correlations carry information on the projected
dark-matter distribution along the lines-of-sight and can thus provide
constraints on cosmology. Considering distant quasars and foreground
galaxies, M\'enard \& Bartelmann (2002) quantified these constraints
and showed that given the large number of parameters involved (the
matter density parameter, $\Omega_0$, the normalisation and shape of
the dark-matter power spectrum, $\sigma_8$ and $\Gamma$, respectively,
and the bias parameter of the galaxies, $b$), the information provided
by the quasar-galaxy correlation function alone is insufficient for
independently constraining all of these parameters.

In this paper, we have investigated what additional information can be
expected from higher-order statistics. Such statistics have similar
weightings of the power spectrum along the line-of-sight and can
measure non-Gaussianities of the density field due to the non-linear
growth of structures. We have specifically considered correlations
beween distant quasars and foreground galaxy \emph{pairs} and showed
that this three-point correlator can be related to the excess scatter
of galaxies around quasars with respect to random positions, which is
a straightforwardly measurable quantity. Using the assumptions that
\begin{itemize}
\item galaxies are linearly biased with respect to the underlying
  dark matter,
\item the dark-matter distribution is described by a CDM power
  spectrum,
\item and the lensing magnifications can be approximated by
$\mu=1+2\kappa$ in the weak lensing regime,
\end{itemize}
we have computed the amplitude of the expected excess scatter of
galaxies in the vicinity of quasars. For a flat cosmology with
$\Omega_0=0.3$ and $\sigma_8=0.93$, we find the amplitude of this
effect to reach the per cent level at angular scales near one arc
minute.

We further showed that combining second- and third-order statistics
allows a pseudo-skewness parameter $S_3'$ to be defined which turns
out to be weakly sensitive to the normalisation and the shape of
the power spectrum. Moreover, if the linear biasing scheme is valid,
this parameter is only sensitive to the matter density $\Omega_0$; the
dependences on the other parameters ($\sigma_8$, $\Gamma$, $\Lambda$
and $\bar b$) are weak or cancel out completely. Thus the skewness
$S_3'$ provides a direct and independent measurement of $\Omega_0$.

We computed the expected angular variation of $S_3'$ and showed that
for a given cosmology and assuming a CDM power spectrum any departure
from the predicted angular shape must be due to a non-linear and/or
stochastic behaviour of the galaxy bias, on scales larger than a few
arc minutes.

Finally, we estimated the signal-to-noise ratio of the expected excess
scatter of galaxies near quasars, which is the main source of noise in
$S_3'$. We derived the noise coming from the finite sampling of
galaxies having realistic distributions. Applying our result to the
Sloan Digital Sky Survey, we find that $S_3'$ should be measurable
on large scales from that survey with about twenty thousand distant and bright 
quasars, allowing thus a direct and independent measurement of $\Omega_0$. 

On smaller angular scales less quasars are required and the 
parameter $S_3'$ can probe the angular range and the 
corresponding physical scales where the linear relation between dark 
matter and galaxy fluctuations may break down.

\section*{Acknowledgments}

We thank Francis Bernardeau, St\'ephane Colombi and Peter Schneider
for helpful discussions, and Ludovic Van Waerbeke for providing his
code on lensing statistics. This work was supported in part by the TMR
Network ``Gravitational Lensing: New Constraints on Cosmology and the
Distribution of Dark Matter'' of the EC under contract
No. ERBFMRX-CT97-0172.

\appendix

\section{Noise estimation}

We use the formalism developed by Szapudi \& Colombi (1996) for
calculating errors due to finite sampling, and we refer the reader to
this paper as well as to the review on ``Large-Scale
Structure of the Universe and Cosmological Perturbation Theory'' by
Bernardeau et al.~(2001). We first define some useful quantities and
then derive the error on our estimator due to Poissonian noise.

\noindent We have the estimator
\begin{equation}
  E^2=\sigma^2\left(\frac{\langle N^2\rangle-\langle N\rangle^2}
                         {\langle N \rangle^2}\right)=
  \sigma^2\left(\frac{\langle N^2\rangle}
                     {\langle N\rangle^2}\right)\;.
\end{equation}
Let $P_N$ denote the probability of finding $N$ galaxies in a cell of
a given size. The factorial moments are defined as
\begin{equation}
  F_k=\langle (N)_k\rangle=\sum(N)_k\,P_N\;,
\label{fk_def}
\end{equation}
where $(N)_k=N(N-1)...(N-k+1)$ is the $k$-th falling factorial of $N$,
and the ensemble average can be evaluated using the known probability
distribution $P_N$. We can write the error on our estimator in terms
of factorial moments,
\begin{eqnarray}
  E^2&=&\sigma^2\left(\frac{F_1+F_2}{F_1^2}\right)\nonumber\\
&=&
   \left\langle\left(\frac{F_1+F_2}{F_1^2}\right)^2\right\rangle_C-
   \left\langle\frac{F_1+F_2}{F_1^2}\right\rangle^2_C\;,
\end{eqnarray}
where the operator $\langle\ldots\rangle_C$ averages over all possible
ways of throwing $C$ cells into the survey volume. In order to compute
the difference, we now perturb $F_1$ and $F_2$ as
$F_1=\bar{F}_1(1+\frac{\delta F_1}{\bar{F}_1})$ and
$F_2=\bar{F}_2(1+\frac{\delta F_2}{\bar{F}_2})$, with $\langle\delta
F_1\rangle=\langle\delta F_2\rangle=0$.

\noindent This yields
\begin{eqnarray}
  E^2&=&\langle\delta F_1^2 \rangle+
    \frac{4\,\bar F_2}  {\bar F_1}  \langle\delta F_1^2\rangle+
    \frac{4\,\bar F_2^2}{\bar F_1^2}\langle\delta F_1^2\rangle-
    \frac{2\,\bar F_2^2}{\bar F_1^2}
    \langle\delta F_1\,\delta F_2\rangle\nonumber\\&-&
    \frac{4\,\bar F_2^3}{\bar F_1^3}
    \langle\delta F_1\,\delta F_2\rangle+
    \frac{4\,\bar F_2^4}{\bar F_1^4}\langle\delta F_2^2\rangle\;.
\label{develop}
\end{eqnarray}
For evaluating this expression, we need to compute the variances and
the covariance of $F_1$ and $F_2$. We will do this using the
generating function of the probability distribution $P_N$ introduced
in Eq.~(\ref{fk_def}),
\begin{equation}
  P(x)=\sum_N\,P_N\,x^N\;.
\end{equation}
The nice property of this quantity is that the factorial moments are
obtained by a Taylor expansion of $P(x)$ around $x=1$. The cosmic
covariance on factorial moments can then be written as
\begin{equation}
  \mathrm{Cov}(F_i,F_j)=
    \left(\frac{\partial}{\partial x}\right)^i
    \left(\frac{\partial}{\partial y}\right)^j\,
  \left.\mathcal{E}(x,y)\right|_{x=y=1}\;,
\end{equation}
and for estimating the finite-sampling error, the error generating
function $\mathcal{E}$ is
\begin{equation}
  \mathcal{E}(x,y)=\frac{P(xy)-P(x)P(y)}{C}
\end{equation}
(Szapudi \& Colombi 1996), where $C$ is the number of cells used,
i.e.~the number of quasars or random positions involved in the
measurement. Thus, we can compute the variances and covariances needed
in Eq.~(\ref{develop}),
\begin{eqnarray}
  \langle\delta F_1^2\rangle&=&\frac{1}{C}\,
  \frac{\partial}{\partial x}
  \frac{\partial}{\partial y}
  \left.\left[P(xy)-P(x)P(y)\right]\right|_{x=y=1}
  \nonumber\\&=&
  \frac{1}{C}\,\left[
    \sum_N N(N-1)\,P_N+N\,P_N-(\sum_N N\,P_N)^2\right]
  \nonumber\\&=&
  \frac{1}{C}\,(F_2+F_1-F_1^2)\;.
\end{eqnarray}
For Poissonian noise, we have the simple relation $F_k=\bar N^k$,
where $\bar N$ is the average number of galaxies in a cell. Thus, the
expression reduces to
\begin{equation}
  \langle\delta F_1^2\rangle=\frac{\bar N}{C}\;.
\end{equation}
In the same way, we obtain
\begin{eqnarray}
  \langle\delta F_2^2\rangle&=&\frac{1}{C}\,
  \frac{\partial^2}{\partial x^2}
  \frac{\partial^2}{\partial y^2}
  \left.\left[P(xy)-P(x)P(y)\right]\right|_{x=y=1}
  \nonumber\\&=&
  \frac{1}{C}\,
  \left[\sum_N N^2(N-1)^2\,P_N-(\sum_N N(N-1)\,P_N)^2\right]
  \nonumber\\&=&
  \frac{1}{C}\,(F_4+4\,F_3+2\,F_2-F_2^2)\nonumber\\&=&
  \frac{4\bar N^3+2\bar N^2}{C} 
\end{eqnarray}
and
\begin{eqnarray}
  \langle\delta F_1\delta F_2\rangle&=&\frac{1}{C}\,
  \frac{\partial}{\partial x}
  \frac{\partial^2}{\partial y^2}
  \left.\left[P(xy)-P(x)P(y)\right]\right|_{x=y=1}\nonumber\\&=&
  \frac{1}{C}\,
  \left[\sum_N N^2(N-1)\,P_N-\sum_N N(N-1)\,P_N\right.
  \nonumber\\&\times&\left.\sum_N N\,P_N\right]\nonumber\\&=&
  \frac{1}{C}\,(F_3+2\,F_2-F_1\,F_2)\nonumber\\&=&
  \frac{2\bar N^2}{C}\;.
\end{eqnarray}

Inserting those terms into Eq.~(\ref{develop}), we find after further
simplifications
\begin{equation}
  E^2\left(\frac{\langle N^2\rangle-\langle N\rangle^2}
                {\langle N\rangle^2}\right)=
  \frac{2}{C\,\bar N^2}+\frac{1}{C\,\bar N^3}\;,
\end{equation}
which is the error due to the finite quasar number.

In order to check this result, we have performed a numerical
simulation in which we measured our estimator on two-dimensional
Poisson distributions of particules. We then computed its
standard deviation for several cell numbers and particular point densities
and find the numerical results in full agreement with the previous
expression. 

The clustering of galaxies can also be taken into account in this calculation. 
To do so, we express the factorial moments of the galaxy number counts in terms 
of n-points cell-averaged correlation functions~:
\begin{equation}
F_k=\bar N_k\,\langle (1+\delta)^k \rangle\,.
\end{equation}
Using the fact that $\langle \delta^2 \rangle=\langle \delta^2 \rangle_c=w(\theta)$, $\langle \delta^3 \rangle=\langle \delta^3 \rangle_c=z(\theta)$ and $\langle \delta^4 \rangle=\langle \delta^4 \rangle_c+3\,\langle \delta^2 \rangle^2=w_4(\theta)+3\,w(\theta)^2$ where $\bar w$, $\bar z$ and $\bar w_4$ are the 2-, 3- and 4-point cell-averaged correlation functions of the galaxies and the subsript $c$ denotes the connected contribution, we can use the expression of the factorial moments in terms of correlation functions in the error expression (Eq. \ref{develop}). One finds~:
\begin{eqnarray}
E^2\left(\frac{\langle N^2\rangle-\langle N\rangle^2}{\langle N\rangle^2}\right)&=&
  \frac{1}{C\,\bar N^3}+
  \frac{2+3\,\bar w}{C\,\bar N^2}+
  \frac{4\,\bar w+2\,\bar z}{C\,N}\nonumber\\
&+&  \bar w_4-4\,\bar z\,\bar w+2\,\bar w^2+4\,\bar w^3\,.
\end{eqnarray}
Under the hierarchical assumption, higher-order moments can be computed by using the skewness and kurtosis parameters of the galaxy distribution~: $\bar z=s_3\,\bar w^2$ and $\bar w_4=s_4\,\bar w^3$. The previous expression can therefore be expressed in terms of $w(\theta)$ only.
Then, each term can be evaluated by considering a power-law correlation function~: $w(\theta)=\theta_0\left( \frac{\theta}{1\,\mathrm{deg}} \right)^{-\gamma}$ and using SDSS measurements of galaxies with $20<r<21$ in order to estimate each parameters. We have~: $\theta_0=10^{-2.3}$ and $\gamma=0.7$ (Connolly et al. 2002) and for the higher-order moments of the same galaxies~: $s_3=4.2\pm0.4$ and $s_4=41\pm10$ (Szapudi et al. 2002).


\begin{thebibliography}{99}

\bibitem[Bacon, Refregier \& Ellis 2000]{BRE00}
Bacon, D., R\'efr\'egier, A., Ellis, R., 2000, MNRAS, 318, 625

\bibitem[Bacon et al. 2002]{Bacon02}
Bacon, D., Massey, R., R\'efr\'egier, A., Ellis, R., 2002, astro-ph/0203134

\bibitem{1} Bartelmann, M., 1995, A\&A, 298, 661

\bibitem{2} Bartelmann, M. \& Schneider, P., 2001, Physics Reports
340, 291

\bibitem{3} Bernardeau, F., Colombi, S., Gaztanaga, E., Scoccimarro,
R., 2002, Physics Reports

\bibitem[Bernardeau, Mellier \& Van Waerbeke 2002]{BMV02}
 Bernardeau, F., Mellier, Y.,  Van Waerbeke, L., 2002, astro-ph/0201032

\bibitem[Bernardeau, Van Waerbeke \& Mellier 1997]{BVM97}
 Bernardeau, F., Van Waerbeke, L., Mellier, Y., 1997, A\&A, 322, 1

\bibitem[Bernardeau, Van Waerbeke \& Mellier 2002]{BVM02}
 Bernardeau, F., Van Waerbeke, L., Mellier, Y., 2002, astro-ph/0201029

\bibitem{4} Blandford, R. D., Narayan, R., 1992, ARA\&A, 30, 311

\bibitem{5} Connolly, A., Scranton, R., Johnston, D. and the SDSS collaboration, astro-ph/0107417, ApJ submitted.

\bibitem{6} Dekel, A., Lahav, O., 1999, ApJ, 520, 24

\bibitem{7} Dolag, K. \& Bartelmann, M., 1997, MNRAS 291, 446

\bibitem{8} Eke, V. R., Cole, S., Frenk, C. S., 1996, MNRAS 282,
263

\bibitem{9} Gunn, J. E., 1967, ApJ, 150, 737

\bibitem[Haemmerle et al. 2002]{Haemm02}
  Haemmerle, H.,  J.-M. Miralles, J.-M., Schneider, P., 
 Erben, T., Fosbury, R.~A.~E.,  
 Freudling, W., Pirzkal, N., Jain, B.,  S. D. M. White, S.~D.~M., 
2002, A\&A, 385, 743

\bibitem[Hoekstra et al.  2002]{Hoekstra02}
  Hoekstra, H., Yee, H.~K.~C., Gladders, M.~D.,  
  Barrientos, L.~F., Hall, P.~B., Infante, L., 2002, \apj in press, 
  astro-ph/0202285

\bibitem[Jain \& Seljak 1997]{BhuvSelj97}
  Jain, B.,  Seljak, U., 1997, \apj, 484, 560

\bibitem[Kaiser, Wilson \& Luppino 2000]{Kaiser00}
  Kaiser, N., Wilson, G., Luppino, G.~A., 2000, astro-ph/0003338

\bibitem[Maoli et al. 2001]{Maoli01} 
  Maoli, R., Van Waerbeke, L., 
  Mellier, Y., Schneider, P., Jain, B., Bernardeau, F., Erbe, T., 
  Fort, B., 2001, A\&A 368, 766

\bibitem{10} M\'enard, B., Bartelmann, M., 2002, A\&A, 386, 784

\bibitem{11} M\'enard, B., Hamana, T., Bartelmann, M., Yoshida, N., 2002, in preparation

\bibitem{12} Peacock, J.A, Dodds, S.J., 1996, MNRAS 280, L19

\bibitem{13} Fry, J. N., Peebles, P. J. E., 1980, ApJ, 238, 785

\bibitem{14} Pen, U.-L., 1998, ApJ, 504, 601

\bibitem[R\'efr\'egier, Rhodes \& Groth 2002]{Ref02}
  R\'efr\'egier, A., Rhodes, J., Groth, E., 2002, astro-ph/0203131

\bibitem[Rhodes, R\'efr\'egier \& Groth 2001]{Rhodes01} 
  Rhodes, J., R\'efr\'egier, A.,  Groth, E., 2001, \apj, 552, L85

\bibitem{15} Schneider, D.,P., Richards, G. T., Fan, X., Hall, P. B.,
Strauss, M. A., Vanden Berk, D. E., Gunn, J. E., Newberg, H. J.,
Reichard, T. A., Stoughton, C., Voges, W., Yanny, B., and the SDSS
collaboration, 2002, AJ, 123, 567

\bibitem{16} Schneider, P., van Waerbeke, L., Jain, B., Kruse, G., 1998,
MNRAS 296, 873

\bibitem{17} Scoccimarro, R., Couchman, H. M. P., 2001, MNRAS 325,1312

\bibitem[Scoccimarro \& Friemann 1999]{Scocci99}
  Scoccimarro, R., Friemann, J., 1999, \mnras 520, 35

\bibitem{18} Scranton, R., Johnston, D., Dodelson, S., Frieman,
J. A., Connolly, A., Eisenstein, D. J., Gunn, J. E., Hui, L., Jain,
B., Kent, S., Loveday, J., Narayanan, V., Nichol, R. C., O'Connell,
L., Scoccimarro, R., Sheth, R. K., Stebbins, A., Strauss, M. A.,
Szalay, A. S., Szapudi, I.,Tegmark, M., Vogeley, M., Zehavi, I. \& the
SDSS Collaboration, 2001, astro-ph/0107416

\bibitem{19} Somerville, R.S., Lemson G., Sigad, Y., Dekel, A.,
Kauffmann, G., White, S. D. M., 2001, MNRAS submitted

\bibitem{20} Szapudi, I., Colombi, S., 1996, ApJ 470, 131

\bibitem{21} Szapudi, I., Frieman, J. A., Scoccimarro, R., Szalay, A. S., and the SDSS collaboration, 2002, ApJ 570, 75

\bibitem[Van Waerbeke et al. 2000]{vW00} 
  Van Waerbeke, L. et al., 2000, A\&A, 358, 30 

\bibitem[Van Waerbeke et al. 2001a]{vW01a} 
  Van Waerbeke, L. et al., 2001a, A\&A, 374, 757

\bibitem{22} Van Waerbeke, L., Hamana, T., Scoccimarro, R., Colombi, S.,
 2001, MNRAS 322, 918

\bibitem[Van Waerbeke et al. 2002]{vW02} 
  Van Waerbeke, L., Mellier, Y., Pell\'o, R., Pen, U-L., 
McCracken, H.~J., Jain, B., 2002, astro-ph/0202503

\bibitem{24}  Verde, L., Heavens, A. F., Percival, W. J., Matarrese, S., and the 2dF collaboration, 2002, MNRAS, in press

\bibitem[Wittman et al. 2000]{Wittman00} 
  Wittman, D.M., Tyson, J.A., Kirkman, D., Dell'Antonio, I.,  
  Bernstein, G., 2000, Nature, 405, 143

\bibitem{25} York, D. G., Adelman, J., Anderson, J. E., Anderson,
S. F., et al. 2000, AJ, 120, 1607


\end{thebibliography}
\end{document}